# Microstructured organic cavities with high-reflective flat reflectors fabricated by using a nanoimprint-bonding process

Takuya Enna[1,†], Yuji Adachi[1,†], Tsukasa Hirao[1], Shun Takahashi[1], Yohei Yamamoto[2], and Kenichi Yamashita[1,*]

[1]*Faculty of Electrical Engineering and Electronics, Kyoto Institute of Technology, Matsugasaki, Sakyo-ku, Kyoto 606-8585, Japan*

[2]*Department of Materials Science and Tsukuba Research Center for Energy Materials Science (TREMS), Institute of Pure and Applied Sciences, University of Tsukuba, 1-1-1 Tennodai, Tsukuba, Ibaraki 305-8573, Japan*

[†]These authors equally contributed.

*e-mail: yamasita@kit.ac.jp

The integration of photonic microstructure into organic microcavities represents an effective strategy for manipulating eigenstates of cavity or polariton modes. However, well-established fabrication processes for microstructured organic microcavities are still lacking. In this study, we propose a nanoimprint-bonding process as a novel fabrication method for microstructured organic microcavities. This process relies on a UV nanoimprint technique utilizing two different photopolymer resins, enabling the independent fabrication of highly reflective reflectors and photonic microstructures without compromising the accuracy of each. The resulting organic microcavities demonstrate spatially localized photonic modes within dot structures and their nonlinear responses on the pumping fluence. Furthermore, a highly precise photonic band is confirmed within a honeycomb lattice structure, which is owing to the high quality factor of the cavity achievable with the nanoimprint-bonding process. Additionally, a topological edge state is also observable within a zigzag lattice structure. These results highlight the significant potential of our fabrication method for advancing organic-based photonic devices, including lasers and polariton devices.

## Introduction

The implementation of optical microcavities in optoelectronic devices offers a viable approach for manipulating light-matter interactions and for regulating the optical functionality of the devices.[1] In particular, microcavities find applications in various types of organic-based devices, such as organic light-emitting diodes,[2–4] solar cells,[5,6] lasers,[7–9] and sensors.[10,11] Vertical-cavity surface emitting laser (VCSEL) structure is one of the typical types of solid-state planar microcavity,[12] in which an optically active layer with a thickness as small as the wavelength of light is sandwiched between a pair of high-reflective mirrors. The VCSEL structure, due to its ability to confine photons strongly and enhance the light-matter interactions, provides an excellent platform for studying the light-matter hybridized state, i.e. polariton states.[1,13,14]

In recent years, substantial research efforts have been dedicated to the integration of photonic microstructures within planar microcavities utilizing organic or organic-inorganic halide perovskite materials.[15–20] This endeavor aims to exert precise manipulation of wavefunctions and energy eigenstates for photonic or polaritonic states within both real and wavenumber spaces. To artificially integrate photonic microstructures within microcavities, the most widely employed technique has been the combination of electron beam lithography followed by reactive ion etching processes.[15,21,22] Although this approach affords exceptional precision in the fabrication of photonic microstructures, its impracticality for mass production arises due to the significant processing time involved. For organic VCSELs, alternative fabrication methods such as ion-beam milling and direct laser patterning have been proposed,[16–18,23,24] exploiting the soft material properties of organic compounds. However, the challenges of the depositing high-reflective distributed Bragg reflectors (DBRs) on non-planar surfaces sometimes limit the accuracy attainable in their fabrication.[18] Nevertheless, those studies have successfully demonstrated the various manipulation schemes of cavity or polariton modes including the localization of photonic or polaritonic wavefunctions within dot structures and the formation of photonic or polaritonic bands within periodic lattice structures. These achievements highlight the need for development of a more feasible fabrication method for microcavities integrated with photonic microstructures, while maintaining precision, accuracy, and reproducibility.

In this study, we present a novel fabrication method for organic microcavities integrated with photonic microstructures, employing nanoimprint-bonding (NIB) process. This method relies on UV nanoimprint lithography technique,[25–27] utilizing two different photopolymer resins. Our approach offers distinct advantages, encompassing microfabrication precision in the in-plane direction of microcavity and the successful implementation of highly reflective mirrors in the vertical dimension. Through detailed angular-resolved photoluminescence (ARPL) spectroscopy for microcavity devices with dot structures, we demonstrate spatial localizations of photonic wavefunctions as well as nonlinear photoluminescence (PL) responses to pumping fluence. Furthermore, we reveal the emergence of highly precise photonic band from an organic microcavity with a honeycomb lattice structure, which is due to the high quality factor achievable with the NIB process. Additionally, by employing a strategic arrangement of these lattice structures, we have observed a topological edge mode in the photonic band dispersion. These observations underscore the large potential of our NIB scheme in advancing the development of organic laser and polariton devices operated at room temperature.[14]

## Results and discussions

For fabricating microstructures into microcavity active layers, direct nanoimprinting methods have been previously demonstrated in lead halide perovskite microcavities.[20,28–30] In organic microcavities, on the other hand, alternative approach for fabricating the microstructure have been proposed including a laser pattering.[18] However, in these approaches, although microstructures can be fabricated with high precision, the subsequent deposition of dielectric thin-film multilayers, acting as the upper DBR, onto a non-planar surface may potentially compromise the quality factor ($Q \sim 500$).[18] On the other hand, an alternative method is the post-processing of microstructures for a planar VCSEL-type microcavity, which can be achieved through the electron-beam lithography and the subsequent reactive ion etching processes.[15,31–33] Meanwhile, a hemispheric microcavity using a microstructured DBR has been reported by taking advantage of mechanically soft nature of organic materials.[17] Nonetheless, these approaches require a substantial number of sequential steps. In contrast, the NIB process we propose herein offers a combination of simplicity in microstructure fabrication and the integration of high-reflective DBRs (see Fig. 1).

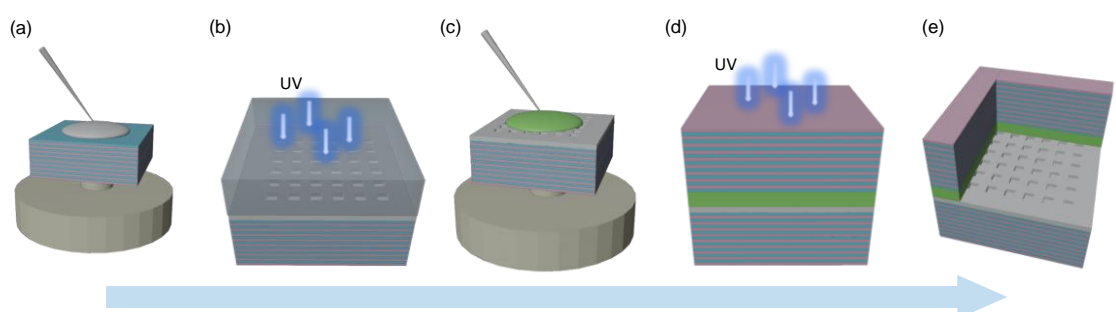

**Figure 1.** Schematics of NIL fabrication process of organic microcavity devices. (a) Spin-coating of low-index photopolymer resin on a DBR substrate. (b) Application of a nanoimprint mold with microstructures and UV exposure. (c) Spin-coating of high-index photopolymer resin doped with organic dye. (d) Bonding of the other DBR substrate with UV exposure. (e) Completed organic microcavity device with photonic structures.

The NIB process employs a pair of silica substrates, onto which high-reflective DBRs are pre-deposited through rf-magnetron sputtering (see Method section and Fig. S1 in the Supplementary Information), along with two types of photopolymer resins. First, we spin-coat one of the photopolymer resins, possessing a refractive index of approximately 1.48 (polymer-1-layer) after the photopolymerization, onto one of the DBR substrates. Subsequently, the spin-coated film is brought into contact with a transparent silica nanoimprint mold at a certain pressure (~35 kPa) and exposed to UV light from the side of the mold. This process yields a polymer film with an attentively engineered photonic structure, as found in Fig. S2 in the Supplementary Information. Following this, the other photopolymer resin, possessing a refractive index of ~1.71, is spin-coated onto the micro-structured polymer-1-layer. The spin-coated film is cured through UV exposure, employing a method similar to the first step, but utilizing another DBR substrate instead of the microstructured imprint mold. In the second polymer layer, an organic luminescent molecule, bis(N,N-di-p-tolylamino-p-

styryl)benzene (referred to as DADSB), is dissolved, forming the polymer-2-layer with an optically active functionality. Absorption and emission spectra of DADSB-doped polymer film is exhibited in Fig. S3 in the Supplementary Information, Consequently, we achieve a microcavity device featuring a microstructured profile in the refractive index distribution along the in-plane direction within the active layer. A thickness of the cavity layer, consisting of the polymeric bilayer, can be controlled within a range of ~2.0 to 2.5 μm through the adjustment of compression pressure during the imprinting process. Notably, the independent fabrications of both the top and bottom DBRs onto the flat silica plates ensure the high reflectivities as the cavity mirrors, leading to the high quality factor of resultant microcavity devices, even when the photonic structures are formed in the cavity layer. This approach enables precise DBR fabrication and simplifies the microfabrication process, thus eliminating the challenges encountered in previous methods.

In the followings, we will show some typical examples of $k$-space dispersion characteristic for organic microcavities fabricated with the NIB process. Figure 2 presents ARPL results for microcavities having isolated square-shaped photonic dot structures with in-plane dimensions of $2 \times 2$ μm$^2$ to $10 \times 10$ μm$^2$. In the dot regions, the thickness of the high-index polymer-2 layer is thicker compared to the surrounding areas, while the thickness of low-index polymer-1 layer is reduced. Therefore, the optical path length within the cavity is longer in the dot regions than the surrounding areas, resulting in the lowering of cavity mode energies. Indeed, the experimental results precisely illustrate the anticipated characteristics of localized wavefunction; the energetically discrete modes emerge in the in-plane $k$ space (denoted as $k_y$ in the figures), and their mode spacing enlarges as the dot size decreases. Each mode exhibits a distribution in the $k_y$ space, signifying the localization characteristics of wavefunction. The wider $k_y$-distribution in smaller-sized dots indicates the stronger localization effects. A straightforward analysis employing the Schrödinger equation for a finite well potential indicates that the energy depth at the dot region is approximately 20 meV (see Supplemental text S1 in the Supplementary Information).

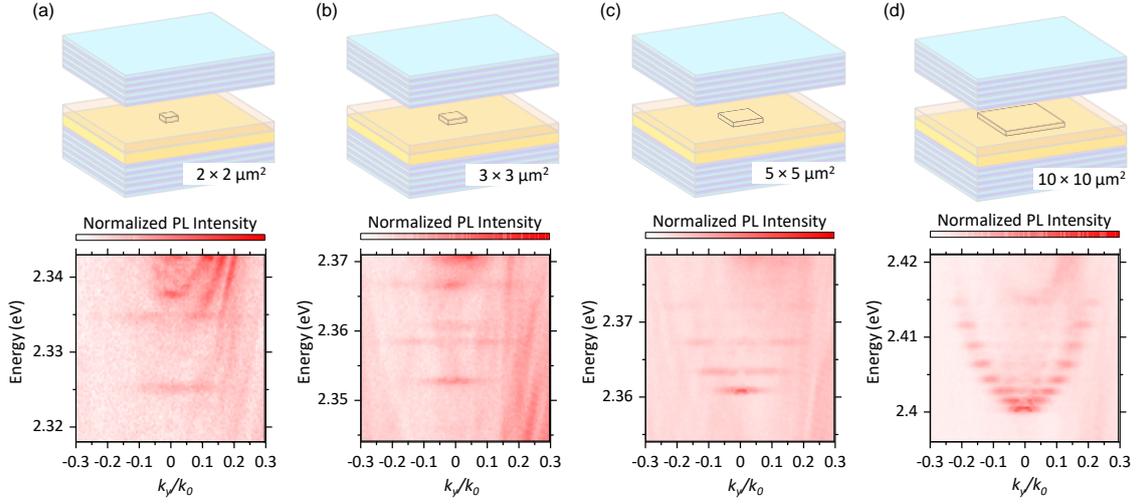

**Figure 2.** ARPL measurements of organic microcavities with dot structures. In-plane dimensions of dot structures are $2 \times 2$ µm² (a), $3 \times 3$ µm² (b), $5 \times 5$ µm² (c), and $10 \times 10$ µm² (d). Upper panels illustrate dot structures inside microcavities, revealing the thickness of DADSB-doped high-index polymer-2 layer is increased at the dot positions. Bottom panels show colormaps of ARPL results along $k_y$ direction. $k_0$ is wavenumber of light in air. The PL intensity is normalized for each colormap.

The nonlinear characteristics under optical pulsed pumping have also been investigated. Figures 3(a) – 3(d) display the colormaps of ARPL results for the device with a $5 \times 5$ µm² dot structure as the pump intensities ($P_{pump}$) increase. Just above 20 µJ, we observe an enhancement in PL intensity at the lowest and second lowest energy levels [at ~2.503 and 2.504 eV, see Figs. 3(a) and 3(b)], indicating the onset of lasing. With further increase in $P_{pump}$, the energetic distribution of the PL signal consolidates only to the lowest energy level [see Figs. 3(c) and 3(d)], suggesting selective enhancement of population in the lowest level. A similar $P_{pump}$-dependent PL distribution is observed also in the $3 \times 3$ µm² dot structure (see Fig. S4 in the Supplementary Information). Figure 3(e) reveals S-shaped profiles in $P_{pump}$-dependent PL intensities for the $3 \times 3$, $5 \times 5$, and $10 \times 10$ µm² dot structures, confirming the occurrence of lasing action. Furthermore, we observe that the smaller-sized dots exhibit more suppressed slope efficiencies within the nonlinearity ranges. This phenomenon is attributed to the Purcell effect, which is more pronounced in smaller-sized dots with lower mode densities.[34] As a result, the effective lasing threshold is reduced from ~30 μJ for the 3 × 3 μm$^2$ dot to ~15 μJ for the 10 × 10 μm$^2$ dot. These experimental results collectively demonstrate the applicability of the NIB technique in the context of organic lasing devices.

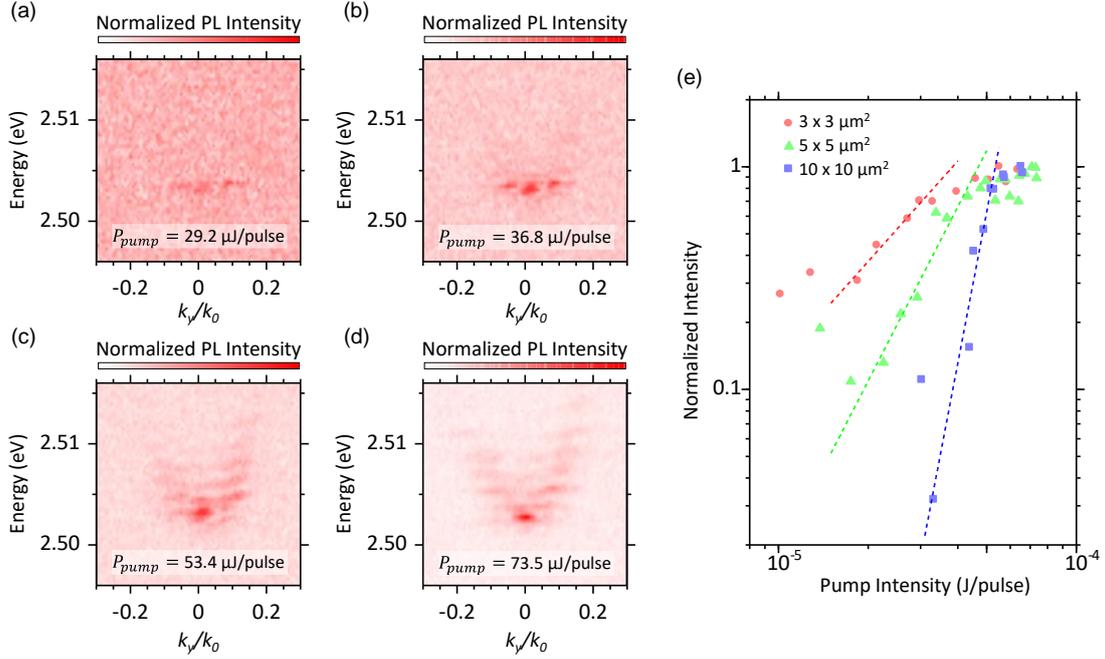

**Figure 3.** ARPL results of 5 × 5 μm$^2$ dot structure under pulsed pumping. (a – d) Colormaps at $P_{pump}$ of 29.2 (a), 36.8 (b), 53.4 (c), and 73.5 μJ/pulse. (e) Pump intensity dependence of PL intensities from 3 × 3 (red), 5 × 5 (green), and 10 × 10 μm$^2$ (blue) dot structures.

Next we present experimental results for a microcavity having a lattice structure.[31,32] The structure examined in this study is a honeycomb lattice with a lattice constant $a$ = 1.28 μm, where circular dots (1.6 μm in diameter) are aligned as a two-dimensional lattice with partial overlap [see Fig. 3(a)]. The ARPL result measured along the $k_y$ direction [at $k_x$ = 0, see Fig. 3(b)] reveals a parabolic dispersion curve around the energy minimum of the cavity mode. The result also exhibits energy gaps around 2.358 eV, implying the photonic band formation. The lowest energy band is attributed to the continuous coupling among the *s*-orbital wavefunctions trapped in the photonic dots. A tight-binding calculation for this honeycomb structure demonstrates that the calculated bonding and anti-bonding bands ($\pi$ and $\pi^*$) can explain well the experimental results (see Supplemental text S2 in the Supplementary Information).[31] Likewise, the ARPL result along the $k_x$ direction also exhibits agreement with the calculation [at $k_y = 0$, see Fig. 3(c)]. From these calculations, the coupling parameters between first and second nearest neighboring dots are evaluated to be ~3.1 meV and ~0 meV, respectively. The overall results of this measurement are effectively visualized in Supplementary movie 1, highlighting the distinct appearance of six Dirac points at the vertices of the hexagonal Brillouin zone [see Fig. 3(d)].[31] The successful observation of a clear photonic band with theoretically predicted dispersion characteristics is owing to the high quality factor $Q$ achievable with the NIB process. Figure S5 exhibits experimentally observed $Q$ is as high as ~1,380. This result is already much higher than previous results for organic resonators fabricated by direct laser writing.[18] However, this value is underestimated due to the insufficient spectral resolution of our measurement system (~0.35 nm) and is actually as high as ~4,000, owing to the fact that the high reflectivities of the cavity mirrors (~0.99) are ensured in devices fabricated with the NIB process.

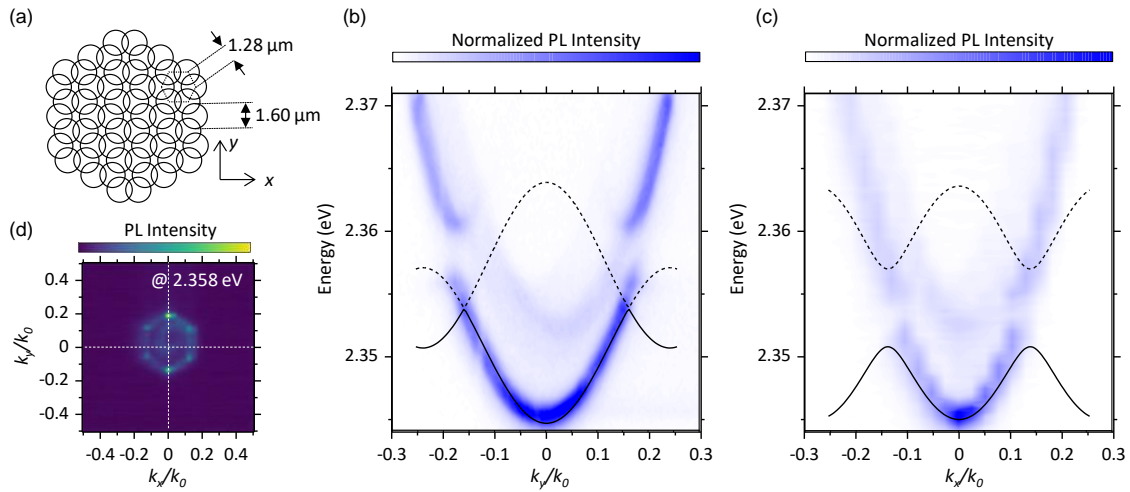

**Figure 3.** ARPL results of honeycomb lattice structure fabricated in organic microcavity. (a) A schematic of fabricated honeycomb lattice structure. (b and c) Colormaps of ARPL measurements along $k_y$ direction at $k_x = 0$ (b) and $k_x$ direction at $k_y = 0$ (c). (d) A colormap of two-dimensional ARPL result at 2.358 eV. Vertical and horizontal white dotted lines reveal cross-section lines in wavenumber space for the colormaps of (b) and (c), respectively.

We also investigate an organic microcavity device with the lattice structure possessing topological properties. An ARPL result for one-dimensional zigzag lattice based on the Su-Schrieffer-Heeger (SSH) model is presented in Fig. S6 in the Supplementary Information.[21,35–37] A flat band is evident within an energy range between *s*- and *p*-bands, indicating the formation of a topological edge mode within the *p*-band. In contrast to the rotationally symmetric *s*-band, the *p*-band is composed of two degenerate modes having wavefunctions distributed oriented in orthogonal directions. In the SSH zigzag lattice, therefore, the coupling strength of *p*-bands is asymmetric between the dots in two different directions. This leads to the emergence of localized modes at one end of the zigzag lattice structure, which unable to couple with neighboring modes, thereby forming a flat band in wavenumber space. Notably, the observation of topological edge modes in organic microcavities has been very few.[23] Our experimental finding confirms the validity of NIB technique to produce the organic topological microcavities.

In the current study, we have not definitively demonstrated that this microcavity system operates within the strongly coupled regime. Consequently, through the above discussions, the modes observed in the dot and lattice structures have been referred to as photonic modes. Nonetheless, the light-matter coupling strength of a planar DBR/polymer-1/DADSB-doped polymer-2/DBR microcavity sample, fabricated using a planar mold without microstructures, can be experimentally assessed, as shown in in Fig. S7 in the Supplementary Information. The ARPL results of this planar microcavity are well explained by the coupled oscillator model with a phenomenological Hamiltonian (see Supplementary text S3 in the Supplementary Information). The coupling strength is estimated to be ~ 200 meV, significantly surpassing the linewidths of the microcavity PL and absorption of DADSB. This result strongly suggests that the microcavity system indeed operates within the strong coupling regime. It is worth noting that further evidence such as changes in the mode curvature at the large wavenumbers and the anti-crossing of lower- and upper-polariton modes, should be investigated as future works. These challenges are very important for establishing a fabrication scheme for room temperature organic polariton devices equipped with photonic microstructures.

## Conclusions

In this study, we introduced the NIB process for fabricating organic microcavities integrated with photonic microstructures. Our NIB approach offers precise control of microfabrication within the in-plane direction of microcavity and incorporates highly reflective mirrors in the vertical dimension. Through detailed ARPL spectroscopy, we demonstrated the spatially localized modes within photonic dot structures. Nonlinear PL responses to pumping fluence was also confirmed, providing evidence of lasing from localized states. Importantly, we observed a clear photonic band with theoretically predicted dispersion characteristics, as the high quality factor is achievable with the NIB process. Moreover, by strategically arranging the lattice structures, we observed a topological edge mode within an SSH zigzag lattice. The successful observations of well-manipulated mode dispersions underscore the significant potential of our NIB fabrication process in advancing the development of organic lasing and also for the room temperature organic polariton devices.

## Materials and methods

### *Microcavity fabrication*

High reflective distributed Bragg reflectors consisting of 9 pairs of $SiO_2/TiO_2$ dielectric multilayers were used as top and bottom mirrors. The reflective bandwidth is 450 – 550 nm. As shown in Fig. 1, a low-index photo-curable resin ($n$ ~1.48, PAK -01-CL, Toyo Gosei) is deposited on the bottom mirror by spin-coating method (accelerated from 500 rpm to 6,000 rpm and 6,000 rom for 60 s). The spin-coated film was imprinted using a transparent silica imprint mold. UV exposure for 180 s (at 365 nm and ~50 mW/cm$^2$) was carried out by using a UV exposure system (M-1S, Mikasa). Microscopic images after releasing the mold are shown in Fig. S1 in the Supplementary Information. Subsequently, a high-index photo-curable resin ($n \sim 1.71$, #18247, NTT-AT) was spin-coated on the microstructured low-index polymer layer. Organic semiconductor molecule, bis(N,N-di-p-tolylamino-p-styryl)benzene (DADSB) was doped into the high-index photo-curable resin with a concentration of 0.4 wt%. The top DBR was put on the spin-coated film and adhered by UV exposure for 300 s. The resultant thicknesses of the low-index and high-index layers are ~0.5 and ~2.0 µm, respectively, at unstructured areas.

*Characterizations*

All optical measurements were performed in atmospheric condition of ~23 °C and ~40 %RH. For steady state PL measurements, a 405-nm cw laser diode (L405P150 & LTC56A/M, Thorlabs) was used for sample excitation. The excitation power was ~150 W/cm$^2$. The PL signal was corrected by an objective with NA ~ 0.5 (UPLFLN 20X, Olympus) and introduced to a CCD spectrometer (Kymera & Newton, Oxford ANDOR). The emission counts were recorded with time-integration of 1 s and accumulation of 10 times. We used a Fourier space imaging setup to obtain the ARPL spectra. The angle limit is estimated from a relationship of $\theta_{\max} = \sin^{-1}(\text{NA})$ to be 30 º.

For above threshold measurements, we used a picosecond laser (PT403, EKSPLA) for sample excitation. The wavelength and pulse width are 355 nm and 15 ps, respectively. The repetition frequency was set at 1 Hz. The pumping intensity was measured with a power meter (PM400, Thorlabs). The emission spectra were measured with a CCD spectrometer (Kymera & Newton, Oxford ANDOR). The emission counts were recorded with time-integration of 1 s and accumulation of 5 times.

## Data availability

The datasets generated during and/or analysed during the current study are available from the corresponding author on reasonable request.

## Acknowledgements

This work is supported by Japan Society for the Promotion of Science, JSPS KAKENHI (Nos. 20KK0088, 22K18794, and 22H00215) and from JST CREST (JPMJCR20T4).

## Conflict of interest

The authors declare no conflict of interests.

## Author contributions

Y.Y. and K.Y. conceived and planned the experiments. H.M. grew BP1T-CN crystals using vapor-phase method. M.N. fabricated dielectric thin-film multilayers using magnetron sputtering. R.O. and T.I. performed ARPL measurements. T.I. and K.Y. performed theoretical calculations. R.O., T.I., and K.Y. analyzed experimental and calculated results. S.T. and K.Y. discussed on topological properties. R.O., T.I., and K.Y. drafted the manuscript and complied figures, with discussion of results and feedback from all authors.